\documentclass[twocolumn,showpacs,showkeys,preprintnumbers,
               amsmath,amssymb,floatfix,10pt]{revtex4}
\usepackage[dvips]{graphicx}
\usepackage{subfigure}
\usepackage{amsmath}
\usepackage{amssymb,amsfonts}
\usepackage{wrapfig}
\usepackage{color}
\usepackage[english]{babel}

\begin{document}



\title{Long-range energy transfer in proteins}

\author{Francesco Piazza}
\affiliation{Ecole Polytechnique F\'ed\'erale de Lausanne,
Laboratoire de Biophysique Statistique, ITP--SB,
BSP-720, CH-1015 Lausanne, Switzerland}
\affiliation{Centre Europ\'een de Calcul Atomique et Mol\'eculaire (CECAM), 
SB CECAM-GE,  PPH 335 station 13, CH-1015 Lausanne, Switzerland.},
\author{Yves-Henri Sanejouand}
\affiliation{
Laboratoire Biotechnologie, Biocatalyse et Bior\'egulation,
UMR 6204 du CNRS, Facult\'e des Sciences et des Techniques,
2, rue de la Houssini\`ere,
44322 Nantes Cedex 3, France.
\email{Yves-Henri.Sanejouand@univ-nantes.fr}}

%
%
%
%
%

\begin{abstract} Proteins are large and complex molecular machines. In order to perform
their function, most of them need energy, {\em e.g.} either in the form of a photon, like
in the case of the visual pigment rhodopsin, or through the breaking of a chemical
bond, as in the presence of
adenosine triphosphate (ATP).
Such energy, in turn, has to be transmitted to specific locations,
often several tens of \AA \  away from where it is initially released.
Here we show, within the framework of a coarse-grained
nonlinear network model,
that energy in a protein can jump
from site to site with high yields, covering in many instances remarkably large distances.
Following single-site excitations, few specific sites are targeted, systematically
within the stiffest regions. Such energy transfers mark the spontaneous formation of
a localized mode of nonlinear origin at the destination site, which acts as an efficient
energy-accumulating centre. Interestingly, yields are found to be optimum for excitation
energies in the range of biologically relevant ones.\end{abstract}

\maketitle

\begin{figure*}[ht!]
\includegraphics[width=17 truecm,clip]{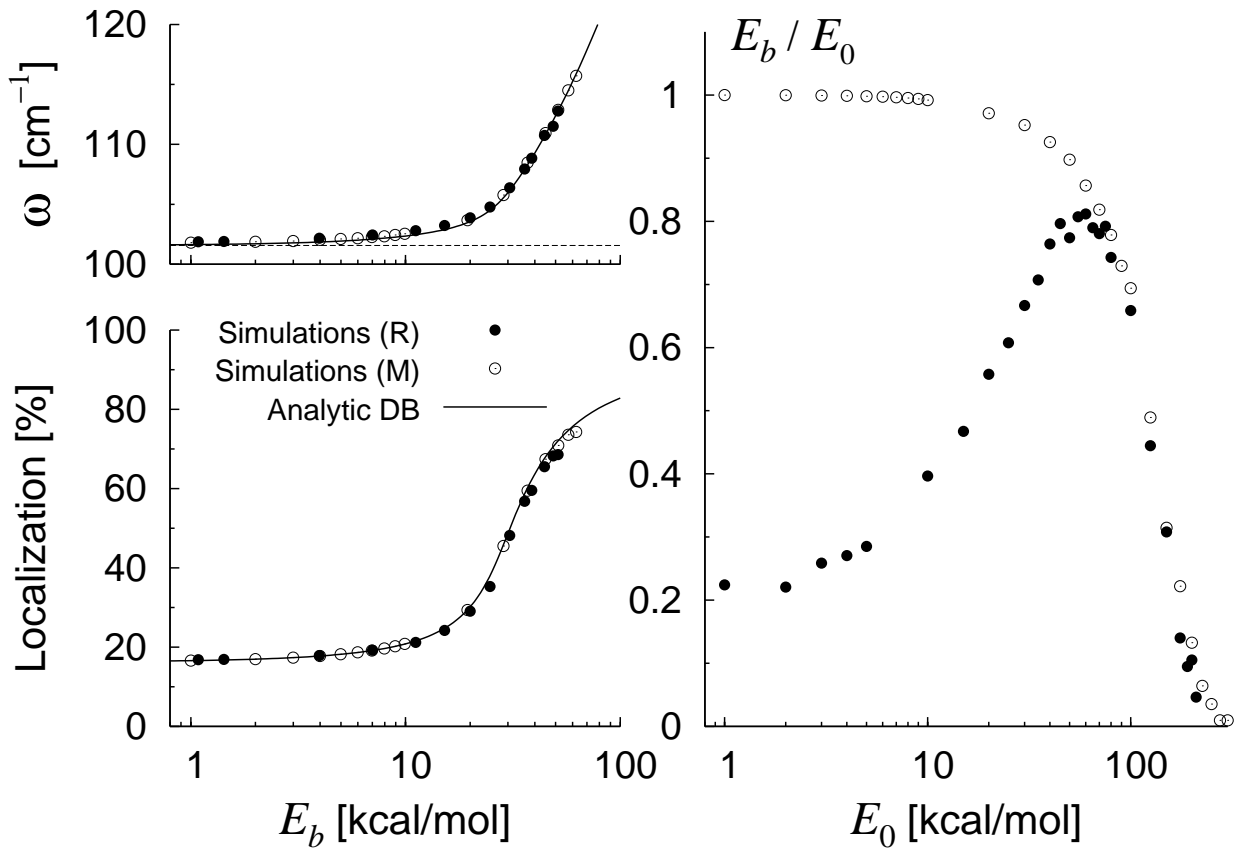}
\vskip 0 mm
\caption{\label{ekept} 
Local kicks cause energy pinning through the excitation of discrete breathers. 
The figure illustrates the formation of a discrete breather at site VAL 177 
in subtilisin (PDB code 1AV7),  a 274 amino-acids enzyme, following the excitation 
of the band-edge normal mode  (M) or a single-site kinetic energy kick at VAL 177 (R), 
the residue where the edge NM is centered. The right panel reports
the energy  $E_{b}$ found in the nonlinear localized mode
as a function of the excitation energy $E_{0}$.
The left panels compare the frequency $\omega$ and the 
localization index $L$ of the nonlinear mode with those of discrete breather 
solutions centered at VAL 177 calculated analytically, as described in 
ref.~\cite{Piazza:08}, showing that 
the nonlinear mode excited after a kick is indeed a discrete breather.
The dashed line in the upper left plot marks the band-edge frequency of 
the protein network ($\omega_{0} = 101.6$ cm$^{-1}$).}
\end{figure*}

Protein dynamics is encoded in their structures and is often critical
for their function~\cite{Karplus:05}. Since the early eighties,
it is well known that vibrational non-harmonicity has to be accounted 
for to understand intra-structure energy 
redistribution~\cite{Levy:82,Kidera:00,Straub:00,Hennig:2002hb,Yamato:06,Leitner:2008rw}. 
Among nonlinear effects, localized modes were suggested to play a key 
role~\cite{Davydov:77,breath-macromol,DB:04}, 
including topological excitations, such as solitons~\cite{Mingaleev:1999pi,dAoOvidio:2005qm} 
as well as Discrete Breathers (DB)~\cite{Aubry:01,Archilla:2002bx}.
The latter, also known as intrinsic localized modes (ILMs),  
are spatially localized, time-periodic vibrations found generically 
in many systems as a combined effect of nonlinearity and spatial 
discreteness~\cite{Flach:2008vx,Sato:2009jt}.
Notably, DBs are able to {\em harvest} from the background and pin down for long times 
amounts of energy much larger than $k_{B}T$.
Indeed, their ability to pump energy from neighboring sites is a 
distinctive signature of DB self-excitation~\cite{Flach:98}, {\it e.g.} observed as a consequence of
surface cooling~\cite{Juanico:07,Livi:2006vn,Piazza:2003cg,Reigada:2001nx} or 
due to modulational instability of band-edge waves 
in nonlinear lattices~\cite{Dauxois:05,Cretegny:1998nx}.
Therefore, provided such 
phenomena are compatible with cellular constraints, it is tempting to
speculate that evolution has found a way to put such long-lived modes at work 
for lowering energy barriers associated with chemical reactions, 
{\em e.g.} for boosting enzyme efficiency during catalytic processes~\cite{Sitnitsky:06}. 

\begin{figure}
\begin{center}
\includegraphics[width=\columnwidth,clip]{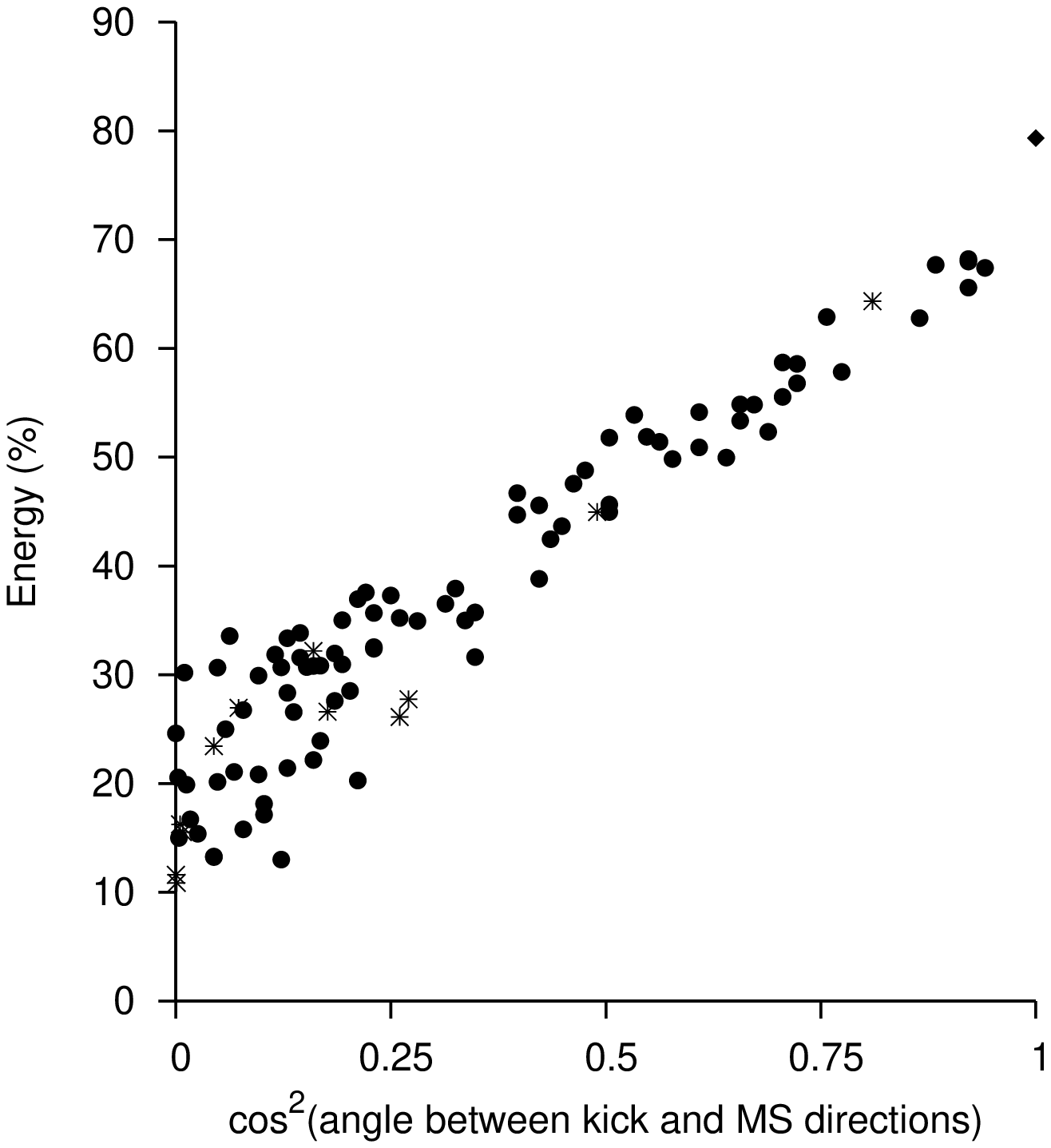}
\end{center}
\vskip -5 mm
\caption{\label{eangle} 
Optimum kick direction for exciting discrete breathers 
in dimeric citrate synthase (PDB code 1IXE).
Percentage of the system energy found in a nonlinear mode as a function
of the direction of the initial kick given to SER 213A, the NM site of the band-edge mode. The latter is measured 
by the angle $\theta$ between the kick direction and the MS unit vector.
In all simulations, the (kinetic) energy of the kick is 55 kcal/mole
and its direction is chosen at random, except when the maximum strain (MS) direction
is picked instead (black diamond at $\cos \theta = 1$).
Filled circles: SER 213A is found to be the most energetic site during the analysis timespan.
Stars: it is another one. In one instance,
while the kick was given in a direction close to the MS direction ($\cos \theta = 0.9$),
the DB jumped on a neighboring site (namely, THR 208A).}
\end{figure}

\section{Nonlinear Network Model}
Recently, within the framework of a  coarse-grained nonlinear network model (NNM), 
we have shown that DBs in proteins feature strongly site-modulated 
properties~\cite{Juanico:07,Piazza:08}. More precisely, we have shown that spatially 
localized band-edge Normal Modes (NM) can be continued from low energies 
to DB solutions centered at the same sites as the corresponding NMs (the NM sites).
Note that the latters lie, as a rule,
within the stiffest regions of a protein~\cite{Juanico:07,Piazza:08}.
More generally, however, DBs display a gap in their excitation spectrum. As a consequence, 
they can ``jump'' to another site as their energy is varied,
following spatial selection rules matching the pattern of DBs localized elsewhere~\cite{Piazza:08}.
As a matter of fact, such jumps realize efficient {\em energy transfers}.
Hereafter, we show that events of this kind, connecting with high yields 
even widely separated locations, can be triggered by
a localized excitation, 
so long as its energy $E_{0}$ lies above a given threshold.

\begin{figure}[t]
\begin{center}
\includegraphics[width=\columnwidth,clip]{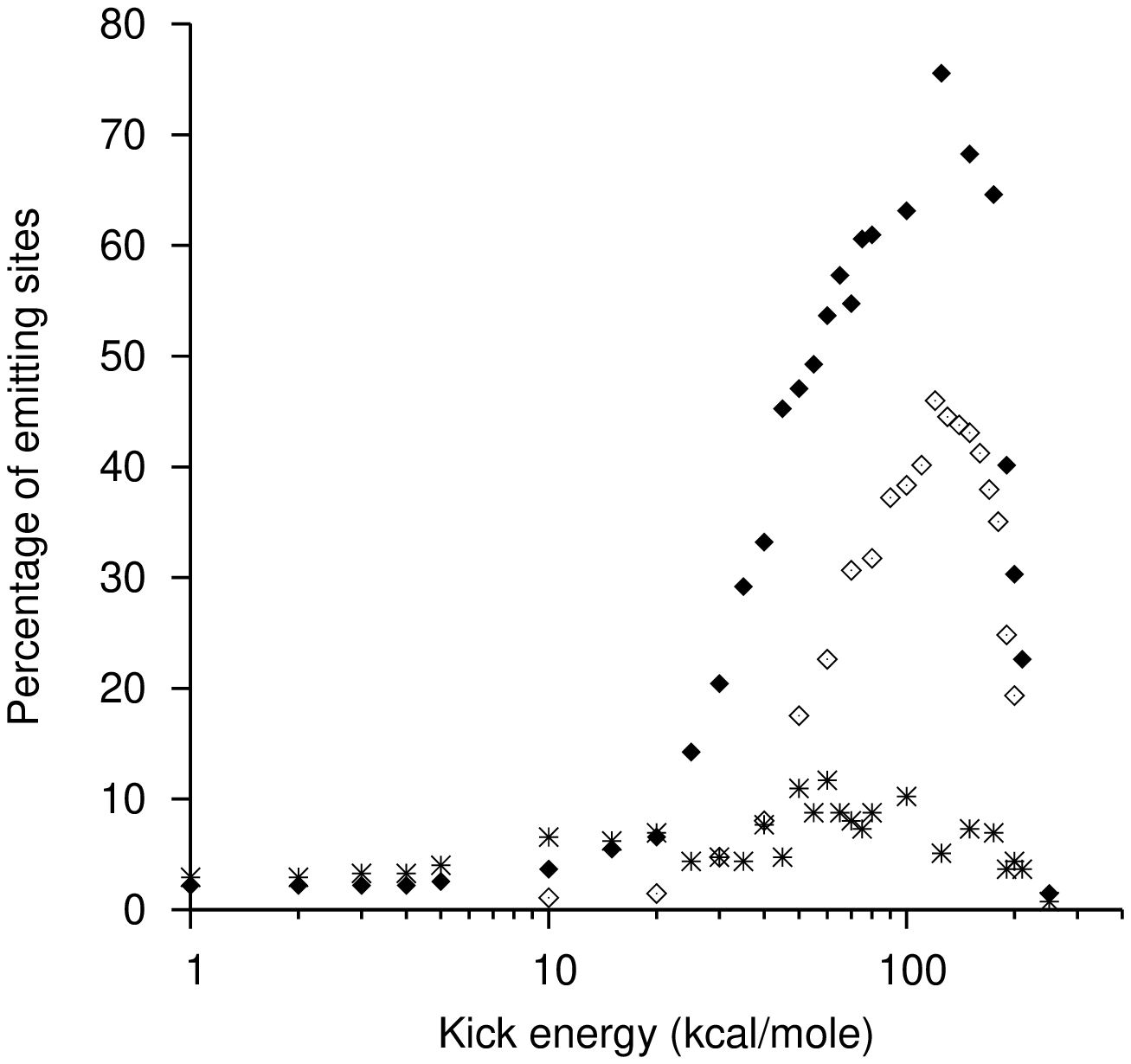}
\end{center}
\vskip -5 mm
\caption{\label{etrans} 
Energy transfer: all-site analysis.
Percentage of sites in subtilisin that transmit most
of the kick energy to the band-edge NM site, VAL 177 (black diamonds), or to the 
NM site of the second edge mode, MET 199 (stars). 
For a given kick energy, each site is kicked once, the most energetic 
nonlinear mode obtained is analyzed, and the site the most involved in this mode is recorded.
When initial excitations are not imparted along the 
local stiffest direction, but are oriented at random, energy transfer towards VAL 177 
is less likely (open diamonds). 
}
\end{figure}

\section{Results}
\subsection{Discrete Breather Excitation}
Fig.~\ref{ekept} summarizes the outcome of one such experiment,
where energy is initially either localized in NM (M) or in real (R) space. 
Typically, the initial excitation is found to spark the formation of
a discrete breather, pinning a variable amount of energy $E_{b}$ at a specific location.
When less than 10 kcal/mole of kinetic energy is injected into the edge NM,
nearly all this energy is kept by the DB, whose overlap with the
edge NM is large at low energies. Increasing $E_{0}$ further,  
the frequency of the excited mode 
detaches from the linear band, while the excitation efficiency $E_{b}/E_{0}$ is eroded.
In fact, as DB localization builds up with energy (see lower left panel), 
the spatial overlap with the edge NM diminishes, thus reducing excitation efficiency~\cite{Kidera:00}.
The same DB is also excited  when the edge NM site
is ``kicked'' along an {\em appropriate} direction, 
namely the maximum stiffness (MS) one~\cite{Piazza:08} 
(see data marked (R) in Fig.~\ref{ekept}).
In this case, however, the excitation becomes more efficient
as $E_{0}$ is increased, since the DB asymptotically approaches a single-site vibration.
For $E_{0} > 100$ kcal/mole, the DB looses its energy, 
which flows rapidly into the system.

\subsection{Directional Specificity}
We find that the maximum strain direction invariably allows for the most efficient  excitation
of a nonlinear mode at a given site. 
Fig.~\ref{eangle} illustrates the efficiency of kicks 
given along the MS direction, with respect to kicks imparted along random directions.
The correlation with the squared cosine of the angle between 
the kick and the MS unit vectors indicates that it is the amount of energy injected
along the MS vector which is the dominant factor allowing for efficient 
excitation of a discrete breather.
\\
\indent
Interestingly, kicking away from the MS direction can promote energy transfer to another site. 
For instance, while a kick along the MS unit vector 
at the NM site of the band-edge mode invariably results in a DB sitting at the same site, 
when the direction of the kick is picked at random
discrete breathers localized elsewhere are also observed (see again Fig.~\ref{eangle}).
In the following, we take advantage of the fact that MS
directions can be easily calculated at any site in any structure~\cite{Piazza:08}
in order to investigate energy transfer in a systematic manner.

\begin{figure}
\begin{center}
\includegraphics[width=\columnwidth,clip]{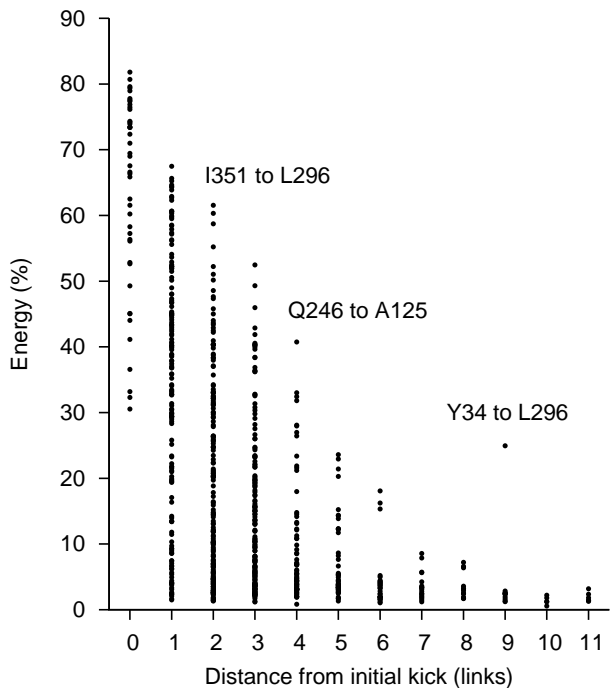}
\end{center}
\vskip -5 mm
\caption{\label{elost} 
Energy transfer as a function of distance from excitation site. 
The figure illustrates the outcome of an all-site 
kick experiment in myosin, a large  746 amino-acids enzyme involved in muscle contraction
(PDB code 1VOM).
The fraction of excitation energy found in the DB
is plotted versus the distance (in units of links in the connectivity graph) 
between the kicked site and the site where the nonlinear mode self-excites. 
The maximum amount of energy found in the DB decreases 
with the number of links separating the feed and the target sites.
For instance, when GLN 246 is kicked, more than 40\% of the energy ends up in a DB 
localized at ALA 125 (the band-edge NM site). This amounts to
four links, corresponding to a span of about 25 \AA \ in real space. 
Otherwise, when a kick is given to ILE 351, GLN 246  or TYR 34, 25-65\% of the 
excitation energy flows either to ALA 125 or LEU 296, 
the NM site of the third edge normal mode.
In cases where more than 30\% 
of the kick energy is transferred away, three sites turn out to be targeted 
half of the times, namely ALA 125 (27\%), LEU 296 (13\%) and GLY 451 (7\%). 
When only long-range energy transfers are considered
(covering three or more links),
the shares raise to 71 \% and 18 \% for ALA 125 and LEU 296, respectively.
In the remaining cases, the DB is found either at LEU 516 (7\%, 14$^{\rm th}$ mode) 
or at ARG 80 (4\%, 10$^{\rm th}$ mode).
}
\end{figure}

\subsection{Energy Transfer}
When a given residue is kicked along the MS direction,
a transfer event can occur when $E_{0} \gtrapprox 20$ kcal/mol (see an example in Fig.~\ref{etrans}). 
At peak transfer, more than 75~\% of such kicks excite a DB localized at the band-edge NM site,
while otherwise energy flows towards the NM site of another edge mode.
Conversely, when the kick is imparted along a random direction, energy transfer is found to be less efficient.
\\
\indent 
Quite generally, a transfer event can be observed when almost any site is kicked, 
and in the majority of cases only a handful of well-defined sites are targeted.
This means that energy transfer can occur between widely separated locations. 
Indeed, as illustrated in Fig.~\ref{elost} for myosin,
only about 5 \% of 55 kcal/mole kicks result in a DB localized at the same location.
For all other kicked sites, a transfer occurs to a DB pinning a decreasing fraction of
the excitation energy, one to eleven links away. Note that all high-yield and long-range energy 
transfers aim at the NM sites of one of the edge NMs, the NM site of the bande-edge mode being
the most likely target. 
Thus, energy systematically flows toward the stiffest regions of the structure.
Interestingly, this
is where functionally relevant residues tend to be 
located~\cite{Juanico:07,Piazza:08,Bahar:05,Lavery:07,Erman:09}. 
\\
\indent 
In one occurrence, more than 20\% of the kick energy ends up in a nonlinear mode localized more
than five links away: following a kick at TYR 34 a remarkable nine-link stretch 
is covered up to LEU 296, making a jump of more than 60 \AA.
However, cases of ultra long-range energy transfer like this are more rare and, at the same time, 
less efficient. In fact, as a consequence of the rather small amount of energy transferred
(nearly 14 kcal/mole), the DB that self-excites at the target site  
is poorly localized (like in Fig.~\ref{ekept}).

\begin{figure}
\begin{center}
\includegraphics[width=\columnwidth,clip]{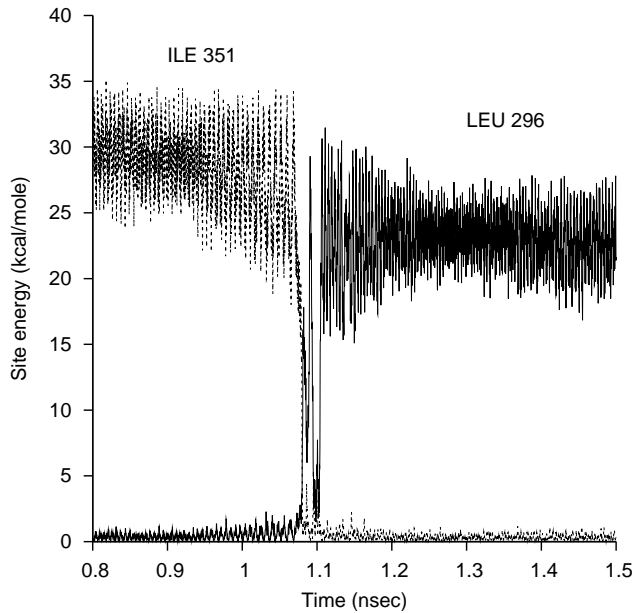}
\end{center}
\vskip -5 mm
\caption{\label{edet} 
Site to site energy transfer in myosin.
The local energies at sites ILE 351 (dotted line) and LEU 296 (solid line)
are plotted as functions of time, after a 55 kcal/mole kick at ILE 351.
The fluctuations occurring well before and after the transfer
reflect the fact that the corresponding nonlinear modes 
are not perfectly localized on both sites.
As a consequence,
they exchange significant amounts of energy with their {\em environs}.
}
\end{figure}

\indent
A more efficient transfer event, covering two links (about 11 \AA), is analyzed in 
Fig.~\ref{edet}. At first, a DB is excited at the kicked site.
However, due to interactions with the background, its energy
slowly but steadily flows into the system. After approximately 1 ns, about 65 \% of
the excitation energy is still there. At  $t=1.1$ ns, this amount of energy 
is rapidly and almost entirely transferred to LEU 296, marking the self-localization of another DB. 
Although the transfer itself is a quite complex process, involving
several intermediate sites, it may well prove to be an example of 
{\em targeted energy transfer}~\cite{Aubry:01}. Indeed, as the energy of the 
DB at the the initial site drops, its frequency diminishes as well. 
This may allow for a transfer to occur if a resonance condition with the frequency of 
another DB is met. The transmission should be irreversible, as a consequence of both DBs' frequency drifts
during energy exchange~\cite{Aubry:01}. Note that, as the energy of the first DB is eroded,
the mode becomes also less and less localized~\cite{Piazza:08}. This, in turn, is likely to increase 
the overlap between the two DB displacement patterns, thus allowing for more efficient energy 
channelling~\cite{Kidera:00,Leitner:2008rw}. 

\begin{figure}
\begin{center}
\includegraphics[width=\columnwidth,clip]{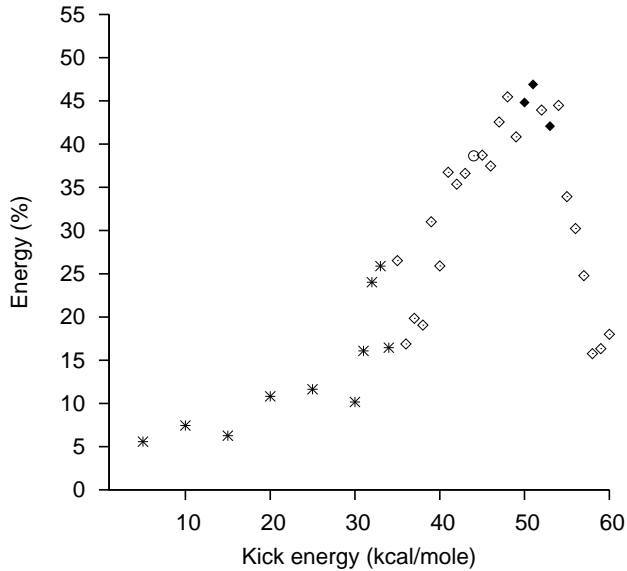}
\end{center}
\vskip -5 mm
\caption{\label{k296} 
Energy transfer in rhodopsin (PDB code 3CAP).
The fraction of energy $E_{b}/E_{0}$ found in the discrete breather
when kicking the site attached to the retinal chromophore (LYS 296) of monomer B
is plotted versus the excitation energy.
Symbols indicate at which site the DB self-localizes: GLU 113 (black diamonds),
CYS 185 (open diamonds), MET 86 (open circle) or another one (stars). 
}
\end{figure}

To gain further understanding on the transfer mechanism, we investigated energy circulation 
in a dimeric form of rhodopsin. Very few high-yield and long-range energy
transfers were recorded between sites belonging to different monomers,
the vast majority of transfer events being confined within the same domain. 
Indeed, in less than 1\% of the instances more than 30\% of the kick energy (55 kcal/mole) 
injected at one monomer is transmitted to the other.
Here, at variance with most protein dimers, the stiffest regions are located 
in monomer bulks, so that the edge NMs are localized far away from the 
interface. This strongly suggests that energy transfers not only target stiff regions, 
but can couple any two sites efficiently only through rather stiff channeling pathways.
On the other hand, when kicking one of the two (almost) equivalent sites of 
rhodopsin that are covalently linked to the retinal chromophore, up to about 
50 \% of the excitation energy ends up in a DB localized at one of 
three specific sites, the targeted location depending upon where (which
monomer) the kick is imparted and on the magnitude of the latter.
Interestingly, Fig.~\ref{k296} reveals that transfer efficiency is
optimum in the narrow range 50-55 kcal/mole, {\em i.e.} exactly the energy of photons
that can be absorbed by the retinal chromophore when it is embedded within
rhodopsin ($\lambda= 500 \div 550$ nm).
Interestingly, the preferentially targeted residue in this energy range (GLU 113)
is known to be involved in the early stages of the signaling cascade following
rhodopsin activation~\cite{Mathies:03}. 

\section{Discussion}
In summary, despite its coarse-grained nature, the NNM framework
is able to provide biologically sensible clues about energy circulation in proteins. 
High-yield and long-range energy transfers systematically pin energy at
the sites the most involved  in a small subset of band-edge linear modes,
that is, within the stiffest  parts of protein structures.
These, in turn, are the regions preferentially hosting residues involved in catalytic 
mechanisms~\cite{Juanico:07,Piazza:08,Bahar:05,Lavery:07,Erman:09}. 
Thus, what our study suggests is that protein structures may have been 
designed, during the course of evolution, so as to allow energy to flow where it 
is needed, {\em e.g.} to, or close to catalytic sites, with the aim of lowering 
the energy barriers that have to be overcome during catalytic processes.

Interestingly, in view of the coarse-grained nature of the NNM scheme, 
the same site-specific, high-yield and long-range energy transfers 
observed in proteins are also likely to occur in other physical systems, possibly 
simpler to engineer and to handle, so long as they share with proteins
both spatial and stiffness heterogeneity.

\section{Methods}
Proteins are modelled as networks of nodes of mass $M$
(the $\alpha$-carbons of their amino-acid residues)
linked by springs. Specifically, in the nonlinear network model 
(NNM)~\cite{Juanico:07,Piazza:08},
the potential energy of a protein, $E_p$, has the following form:
\begin{equation}
\label{FPU}
  E_p=\sum_{d_{ij}^0 < R_c} \left[
                             \frac{k_{2}}{2} (d_{ij}-d_{ij}^0)^2 +
                             \frac{k_{4}}{4} (d_{ij}-d_{ij}^0)^4
                            \right] 
\notag
\end{equation}
where $d_{ij}$ is the distance between particles $i$ and $j$,
$d_{ij}^0$ their distance in the equilibrium
structure (as {\em e.g.} solved through X-ray crystallography)
and $R_c$ is a distance cutoff that specifies which pairs of nodes are interacting.
Note that $k_4=0$ corresponds to the widely used Elastic Network Model 
(ENM)~\cite{Tirion:96,Bahar:97,Hinsen:98}, which has
proven useful for quantitatively describing amino-acid fluctuations
at room temperature~\cite{Tirion:96,Bahar:97,Phillips:07},
as well as for predicting and characterizing large-amplitude functional motions of 
proteins~\cite{Tama:01,Delarue:02,Gerstein:02,Ma:05b,Nicolay:06},
in agreement with all-atom models~\cite{Karplus:76,Marques:95,Perahia:95}, 
paving the way for numerous applications in structural biology~\cite{NMA}, 
such as fitting atomic structures into 
low-resolution electron density maps~\cite{Tama:04,Delarue:04}, 
or providing templates for molecular replacement techniques~\cite{Elnemo1}. 
As in previous NNM studies~\cite{Juanico:07,Piazza:08},
we take $R_c=$10  \AA, $k_4=5$ kcal/mol/\AA$^4$ and
fix $k_2$ so that the low-frequency part of the linear spectrum matches
actual protein frequencies,
as calculated using realistic force fields~\cite{NMA}.
When $M=110$ a.m.u. (the average amino-acid residue mass),
this gives $k_2=5$ kcal/mol/\AA$^2$.

For each site in a given structure, the maximum-stiffness (MS) direction is 
computed through the Sequential Maximum Strain algorithm~\cite{Piazza:08}.
Following an initial kinetic-energy impulse (kick) at a specific site along the local MS
unit vector, a 2-ns microcanonical simulation is performed.
After a 1-ns transient period during which 
a part of the excitation energy flows into 
the system, the velocity-covariance matrix is computed. Its first eigenvector 
provides the pattern of correlated site velocities involved in the dominant (most energetic)
nonlinear mode (the DB). Accordingly, a transfer is recorded to the site at which the 
first principal mode (PM1) is found localized.
Projecting the system trajectory on PM1
yields fair estimates of the DB frequency and average energy~\cite{Juanico:07}. 
The localization index $L$ of a DB centered at site $m$ is obtained from the weight 
of the latter in the normalized 
displacement pattern of the DB, namely $L = 100 \times \sum_{\alpha=x,y,z} [\xi_\alpha(m)]^2$,
where $\xi_\alpha (m)$ are the components at site $m$ 
of PM1.


\end{document}